\documentclass[sn-mathphys]{sn-jnl}

\jyear{2024}%

\theoremstyle{thmstyleone}%
\theoremstyle{thmstyletwo}%

\theoremstyle{thmstylethree}%

\usepackage{lineno}

\graphicspath{ {./images/} }
\usepackage{bm}
\usepackage{amsmath}
\usepackage{subfig}
\usepackage{array}
\newcolumntype{x}[1]{>{\centering\arraybackslash}p{#1}}

\colorlet{tablerowcolor}{gray!10}

\usepackage{csquotes}

\usepackage{subfiles}
\usepackage{lscape}
\usepackage{comment}
\usepackage[export]{adjustbox}
\usepackage{pdfpages}
\usepackage{caption} 
\usepackage{longtable}
\usepackage{t1enc}

\begin{document}

\title[Collective moderation in online discussions]{Collective moderation of hate, toxicity, and extremity in online discussions}


\author*[1,2]{\fnm{Jana} \sur{Lasser}}\email{jana.lasser@uni-graz.at}
\equalcont{These authors contributed equally to this work.}

\author*[1,3]{\fnm{Alina} \sur{Herderich}}\email{alina.herderich@uni-graz.at}
\equalcont{These authors contributed equally to this work.}

\author[4,5]{\fnm{Joshua} \sur{Garland}}

\author[3,6]{\fnm{Segun Taofeek} \sur{Aroyehun}}

\author[2,3,6]{\fnm{David} \sur{Garcia}}

\author[2,5,7]{\fnm{Mirta} \sur{Galesic}}

\affil*[1]{\orgname{IDea\_Lab, University of Graz}, \city{Graz}, \country{Austria}}

\affil[2]{\orgname{Complexity Science Hub Vienna}, \city{Vienna}, \country{Austria}}

\affil*[3]{\orgname{Department of Computer Science and Biomedical Engineering, Graz University of Technology}, \city{Graz}, \country{Austria}}

\affil[4]{\orgname{Center on Narrative, Disinformation, and Strategic Influence, Arizona State University}, \city{Tempe}, \country{AZ, United States}}

\affil[5]{\orgname{Santa Fe Institute}, \city{Santa Fe}, \country{NM, United States}}

\affil[6]{\orgname{Department of Politics and Public Administration, University of Konstanz}, \orgaddress{\city{Konstanz}, \country{Germany}}}

\affil[7]{\orgname{Vermont Complex Systems Center, University of Vermont}, \orgaddress{\city{Burlington},  \country{VT, United States}}}

\abstract{
In the digital age, hate speech poses a threat to the functioning of social media platforms as spaces for public discourse. Top-down approaches to moderate hate speech encounter difficulties due to conflicts with freedom of expression and issues of scalability. Counter speech, a form of collective moderation by citizens, has emerged as a potential remedy. Here, we aim to investigate which counter speech strategies are most effective in reducing the prevalence of hate, toxicity, and extremity on online platforms. We analyze more than 130,000 discussions on German Twitter starting at the peak of the migrant crisis in 2015 and extending over four years. We use human annotation and machine learning classifiers to identify argumentation strategies, ingroup and outgroup references, emotional tone, and different measures of discourse quality. Using matching and time-series analyses we discern the effectiveness of naturally observed counter speech strategies on the micro-level (individual tweet pairs), meso-level (entire discussions) and macro-level (over days). We find that expressing straightforward opinions, even if not factual but devoid of insults, results in the least subsequent hate, toxicity, and extremity over all levels of analyses. This strategy complements currently recommended counter speech strategies and is easy for citizens to engage in. Sarcasm can also be effective in improving discourse quality, especially in the presence of organized extreme groups. Going beyond one-shot analyses on smaller samples prevalent in most prior studies, our findings have implications for the successful management of public online spaces through collective civic moderation.
}

\maketitle


\paragraph{Classification}
Social and Political Sciences > Political Sciences, Psychological and Cognitive Sciences, Social Sciences

\paragraph{Significance Statement}
In the digital age, hate speech poses a threat to the functionality of social media platforms as spaces for public discourse. Given the difficulties with top-down moderation of online platforms, counter speech has emerged as a potential remedy. This work provides evidence for (and in some cases against) various existing, often intuitive recommendations for engagement in counter speech against online hate speech. It provides guidance to citizens that wish to moderate online discourse in organized groups or as individuals and contribute to a healthier exchange of ideas.

\section{Introduction}
\label{sec:intro}

Social media platforms enable networking and information spreading at an unprecedented scale. While these platforms open a host of opportunities for learning, entertainment, and beneficial joint action, they are also plagued by various types of incivility and misinformation. Attempts to regulate these problems top-down, by the companies who run the platforms or by the governments, have been met with mixed success and a lot of distrust from various parts of the general public~\cite{hangartner2021empathy}. It can be useful to view social media platforms as a common-pool resource~\cite{derosnay2020digital} of truthful, respectful, supportive, and entertaining communication, which can be depleted by misuse. It has been shown that common-pool resources can be effectively managed by self-organized local communities~\cite{ostrom1990governing} that can detect and stop misuse. 

Here we analyze the digital traces of self-organized citizen response to one of the most harmful types of misuse of social media platforms: hate speech. Online hate can not only lead to real-world violence~\cite{buerger2021counterspeech,muller2020hashtag}, but it also depletes the quality of communication on social media platforms. After witnessing hate towards themselves or others in their community, people can be reluctant to share their opinions truthfully and can be motivated to respond in kind. This in turn further contributes to the overall toxic atmosphere in which many do not feel supported and might eventually withdraw their participation~\cite{ziegele2016not}.

Past research has suggested that bottom-up citizen-generated responses to hate, also known as ``counter speech'', can increase the overall quality of communication on social media platforms~\cite{buerger2021counterspeech,citron2011intermediaries}, especially when organized~\cite{garland2022impact}.  Counter speech can be seen as a form of bottom-up collective civic moderation~\cite{friess2021collective}, where citizens monitor and maintain the civility of conversations on a social media platform through forms of interaction that the platform commonly provides. As such, counter speech does not rely on formal top-down moderation infrastructure, instead empowering individuals and communities to promote respectful discourse organically. We now ask: What dimensions of discourse can help moderate online conversations marked by hate? Many potentially useful forms of counter speech have been suggested, including warnings of negative consequences, pointing out hypocrisy and contradictions, showing hostility and aggression, or inducing positive emotions through humor \cite{benesch2016considerations,mathew2018analyzing}. Studies using human coding of subsets of discourse ~\cite{friess2021collective,keller2020combatting,mathew2018analyzing,mathew2019thou,stroud2015changing,ziegele2018journalistic,wright2017vectors} and experiments~\cite{alvarez2018normative,corrington2022influence,hangartner2021empathy, munger2017tweetment} have produced important results, showing that following the norms of rationality (providing reasons and evidence), constructiveness (solution-oriented discourse), and politeness \cite{friess2021collective}, appealing to moral principles \cite{munger2017experimentally} and encouraging empathy for the victims \cite{munger2017tweetment, hangartner2021empathy} can lead to a better deliberative quality and less hate in subsequent discourse. However, past studies were limited to relatively small snapshots of online discourse at a single point in time. In addition, controlled experiments on hateful behavior are nearly infeasible while preserving participant safety and ensuring informed consent. To understand the real-world interplay of hate and counter speech, we need to measure different dimensions of discourse in large textual corpora over longer periods of time. 

We conduct extensive analyses of discourse unfolding over four years on Twitter in Germany. We analyze ``discussion trees'' originating from tweets posted by German news organizations, journalists, bloggers, and politicians~\cite{garland2020countering,garland2022impact}. In total, this corpus contains 1,150,469 tweets posted from 2015 to 2018 by 130,548 different users. Besides the general public, two organized citizen groups participated in the Twitter conversations in our corpus. One, Reconquista Germanica (RG), was supportive of the German populist party Alternative für Deutschland (AfD) whose platform is dominated by opposition to immigration and euroscepticism. The other, Reconquista Internet (RI), was a citizen-organized group aiming to counter the anti-immigration narratives promoted by RG (see~\cite{keller2020combatting} for an in-depth description of these two groups). In this political period, marked by a large influx of refugees into Germany which sparked heated discussions and political polarization, some of the tweets from the two organized groups as well as from the general public exhibited hate towards those with opposing views. 

Although specific to German Twitter, our sample is of particular interest for the study of hate- and counter speech. The data on which we base our analysis was collected in a period of controversial political debate including organized hate (RG) and counter speech groups (RI). In addition, posts are associated with prominent media outlets (e.g., ARD, one of Germany's largest public-service broadcasters). This helps us to capture the intended phenomena, which can be seen as a form of theoretical sampling~\cite{glaser_strauss_1967}. Like in other countries, social media has become a central means of gathering political information and debating political events~\cite{medianewssurvey2023, digitalnewsreport2023} in Germany. Despite their differences, similar dynamics have been observed over multiple social media platforms~\cite{avalle2024persistent}. Our sample allows us to study political debate over a long period of time in a large group of individuals in a non-English speaking country and as such is complementary to previous research.

We investigate how hate, defined as insults, discrimination, or intimidation, spreading fearful, negative, and harmful stereotypes, calling for exclusion or segregation, inciting hatred, and encouraging violence against individuals or groups on the grounds of their supposed race, ethnic origin, gender, religion, or political beliefs~\cite{bakalis2015,blaya2019cyberhate,weber2009manual,youtube_guide,twitter_guide,facebook_guide} changes after tweets characterized by different discourse dimensions. To evaluate the robustness of our results, we track three further indicators of \textit{discourse quality}: \textit{toxicity}, defined as rude, disrespectful, or unreasonable language that is likely to make someone leave a discussion~\cite{perspective-website}; as well \textit{extremity of speech} and \textit{extremity of speakers}, defined as speech or speakers resembling organized extreme political groups (RG or RI, respectively, \cite{garland2022impact}; see SI Section S1 for details). 

We measure different discourse dimensions that can contribute to lower or higher discourse quality: different \textit{argumentation strategies} (from providing mere opinions to constructive comments and insults), \textit{ingroup/outgroup references}, and positive or negative flavors of \textit{emotional tone} of preceding discussions. To measure discourse quality and different dimensions of discourse, we use newly developed and pre-existing machine-learning classifiers based on human coding. We analyze the relationship between discourse dimensions and quality within individual reply pairs (micro level), within discussion trees (meso level) and over days (macro level), providing a nuanced picture of the dynamics of discourse at different levels (see Methods and SI Sections S9-S12). While our results build on observational data, our large-scale corpus spanning several years allows us to approximate the measurement of causal effects beyond simple correlation analyses. 

\section{Results}
\label{sec:results}
Using newly developed and pre-existing classifiers based on human judgment (see Methods), we obtained measures of discourse quality and of different dimensions of discourse for each tweet in the 130,127 Twitter conversations sampled from the 1,461 days starting on January 1, 2015 and ending on December 12, 2018 (see Methods and SI Tab.~S2 for details and examples of classified tweets). We first present descriptive results for each measure separately (Fig.~\ref{fig:all_trends}) and then explore the relationships between discourse quality and different dimensions of preceding discourse (Fig.~\ref{fig:ardl_results}). 

\subsection{Trends in dimensions and quality of discourse over time}
\begin{figure}[!ht]
    \centering
    \includegraphics[width=1\textwidth]{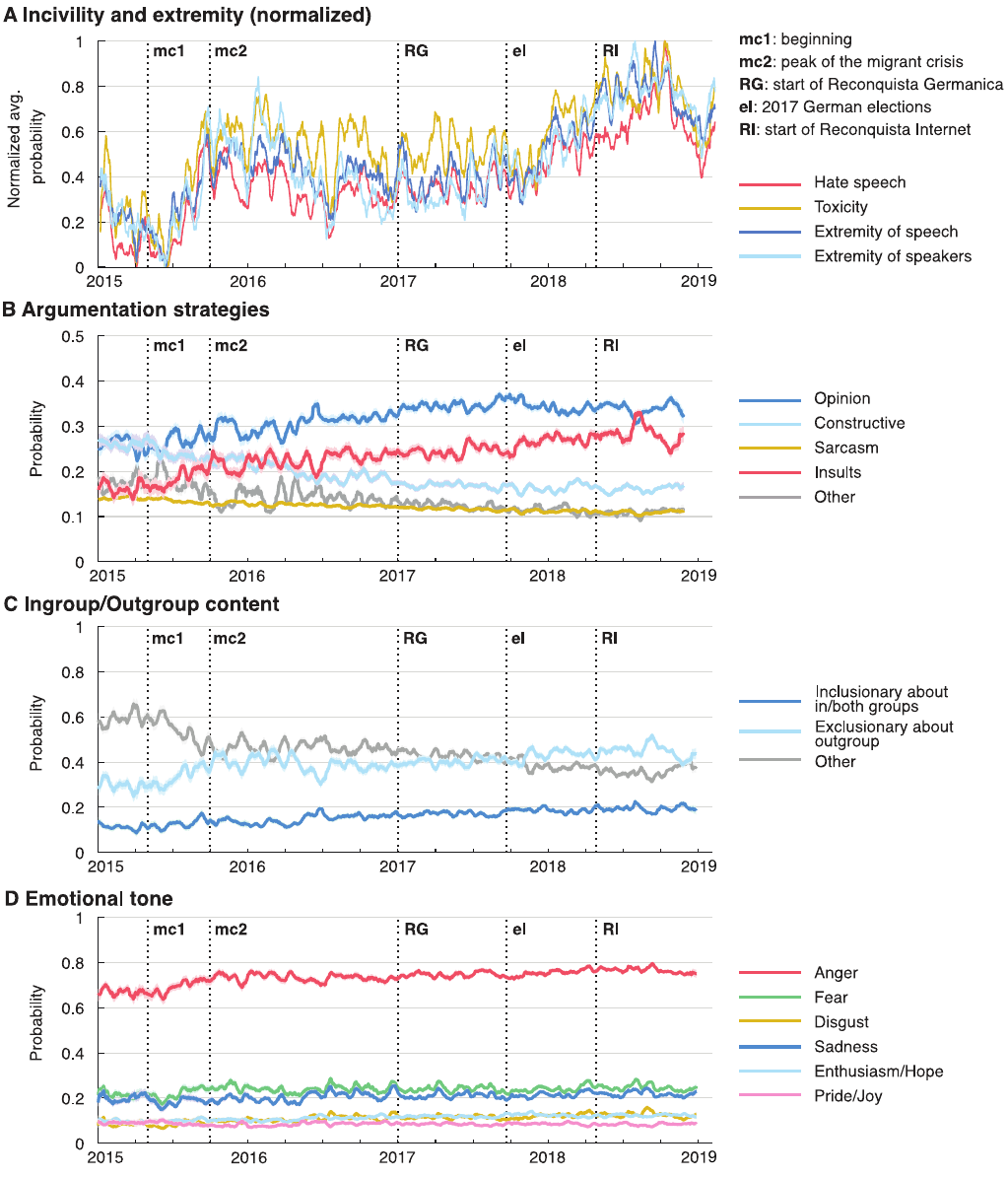}
    \vspace{0.1 cm}
    \caption{\footnotesize\textbf{Measures of discourse quality and dimensions of discourse}. \textbf{(A)} Normalized measures of discourse quality over time, with 0 representing their minimum and 1 their maximum value across the whole data set. Raw values are shown in SI Fig.~S6. \textbf{(B)} Probability of different argumentation strategies over time. \textbf{(C)} Probability of different goals regarding ingroup/outgroup over time. \textbf{(D)} Probability of different emotional tones over time. \textit{Note}. ``Other'' in argumentation strategies refers to ambiguous tweets or tweets that did not fall in any of the other categories. ``Other'' in Ingroup/outgroup content refers to tweets neutral with respect to group identity or tweets where a speaker's identity was not apparent. All measures are on a scale from 0 to 1. For hate speech, toxicity, argumentation strategies, ingroup/outgroup content, and emotional tone, higher values denote a higher probability that a human rater would perceive a tweet as hateful or toxic, or detect a certain strategy, ingroup/outgroup related goal or emotional tone in the tweet. For extremity of speech, higher values denote a higher classifier probability that a tweet is similar to extreme political speech exemplified either by the discourse of Reconquista Internet or of Reconquista Germanica. For the extremity of speakers, higher values denote a higher relative frequency of speakers whose tweets are labeled as containing extreme political speech. Error bands denote standard errors. All trends are smoothed over a two-week window. Thicker vertical lines denote several relevant events: mc1=beginning and mc2=peak of the migrant crisis, RG=start of Reconquista Germanica, el=2017 German elections, RI=start of Reconquista Internet. Additional details are provided in the SI Section~S7.}
    \label{fig:all_trends}
\end{figure}

Panel A of Fig.~\ref{fig:all_trends} shows trends for different measures of discourse quality over time. As each trend has a different range of variation, for easier comparison we normalized them to the scale from their respective minimum and maximum values. This makes it easier to see that all four trends follow similar trajectories, which emphasizes the robustness of our results, as hate speech, toxicity, and extremity measures are each based on different, independently developed classifiers (see Methods for details). A sharp increase in trends - denoting the deterioration of discourse quality on all measures - co-occurs with the 2015 migrant crisis in Germany, which peaked in the fall of that year. The trends then largely remain at the same level for a couple of years, possibly helped by the organized extreme right Twitter group Reconquista Germanica which was established in late 2016 and early 2017. There is another surge in all trends after the German elections in the Fall of 2017 which established the position of the extreme right AfD party (which was supported by Reconquista Germanica) as the third strongest party in the German parliament. Finally, there is a brief dip in all trends after the establishment of the organized counter group Reconquista Internet in late April 2018 and mass real-world protests across the political spectrum occurring in the summer of 2018~\cite{wiki:Chemnitz,wiki:Unteilbar}. Afterwards, the trends start to rise again but do not reach the previous levels - at least not before the end of this time series, although it is possible that the debate surrounding the migrant crisis increases in toxicity with the length of the discussion~\cite{cinelli2021dynamics, avalle2024persistent}. 
Toxicity and hate are not reserved for one or the other political extreme. Speakers and speech from both extremes show higher levels of toxicity and hate than the more neutral speakers and speech (shown in SI Fig.~S7) indicating that the use of hateful language is not reserved for groups from a particular end of the political spectrum~\cite{cinelli2021dynamics}. That said, extreme speech and speakers similar to Reconquista Germanica exhibit higher levels of hate speech and toxicity than those similar to Reconquista Internet. For an in-depth exploration of the extremity trends, see~\cite{garland2022impact}.

\clearpage

Next, Fig.~\ref{fig:all_trends}B shows trends in the use of different argumentation strategies. There is an increasing trend in expressing opinions---not necessarily objective, but without insults, for example \textit{``...most people don’t care what politics are pursued...''}. The use of insults is also on the rise, for example, \textit{``I have to vomit seeing [a politician's] shitface''}. In contrast, the levels of constructive comments have been decreasing slowly throughout the studied period (for example \textit{``According to the Ministry of Interior, under the CDU 83\% of anti-Semitic crimes were committed by right-wing extremists''}, and other comments providing information, asking honest questions, pointing out negative consequences, calling somebody out for behavior or choice of words, or exposing hypocrisy or contradictions). The levels of comments containing sarcasm, irony and cynicism (for example, \textit{``Without being able to compromise he doesn’t belong in politics. He should apply at Karl Lagerfeld.''}) have also been decreasing.

Fig.~\ref{fig:all_trends}C shows that over time the content of discourse becomes more and more about participants' outgroups. The most frequent goal of tweets containing ingroup or outgroup content is the exclusion of outgroups (for example, \textit{``Joining refugee’s families will be the end of Germany as we know it! The end of German culture and way of life!''}). Inclusionary statements about own or both groups, or at least statements treating both groups equally are quite rare (for example, \textit{``I recommend reading our constitution. It holds for all of us.''}), and for the rest of the analyses, we group them together. Both types of statements become more frequent over time, with the ratio of exclusionary and inclusionary statements at roughly 2:1. 

Finally, in Fig.~\ref{fig:all_trends}D we explore the dynamics of tweets' emotional tone. The most prominent result is that anger dominates the emotional signature of the discourse in this corpus, followed by fear and sadness. The increase in anger echoes increases in hate, toxicity, and extremity over time (Fig.~\ref{fig:all_trends}A, as well as the increase in the use of insults, Fig.~\ref{fig:all_trends}B, and exclusionary statements about outgroups, Fig.~\ref{fig:all_trends}C). Furthermore, the four negative emotions mostly correlate moderately with each other, while the four positive emotions show two clear clusters: enthusiasm and hope, and joy and pride (SI Tab.~7). This is partially in line with the results of~\cite{marcus2017measuring} who found strong correlations between enthusiasm, hope, and pride. Our results establish pride as a separate construct, in line with~\cite{sullivan2014collective} who stresses the role of group pride as an important and ubiquitous collective emotion. Given the correlations between emotions, we use the reduced set of four negative and two combined positive emotions in the analyses that follow.

\subsection{Relationships between discourse quality and dimensions of preceding discourse}
We explore the relationship of discourse quality with different argumentation styles, ingroup/outgroup content, and emotional tone (see SI Section S3.6 for example tweets in each category) by conducting statistical analyses on three different levels of discourse: the level of individual reply pairs (micro), discussion trees (meso), and days (macro level) (see Methods and SI Section S9 for details). Our analyses explore the relationship of each dimension of discourse on the overall quality of discourse, while controlling for all other dimensions of discourse.

\begin{figure}[!ht]
    \centering
    \includegraphics[width=\textwidth]{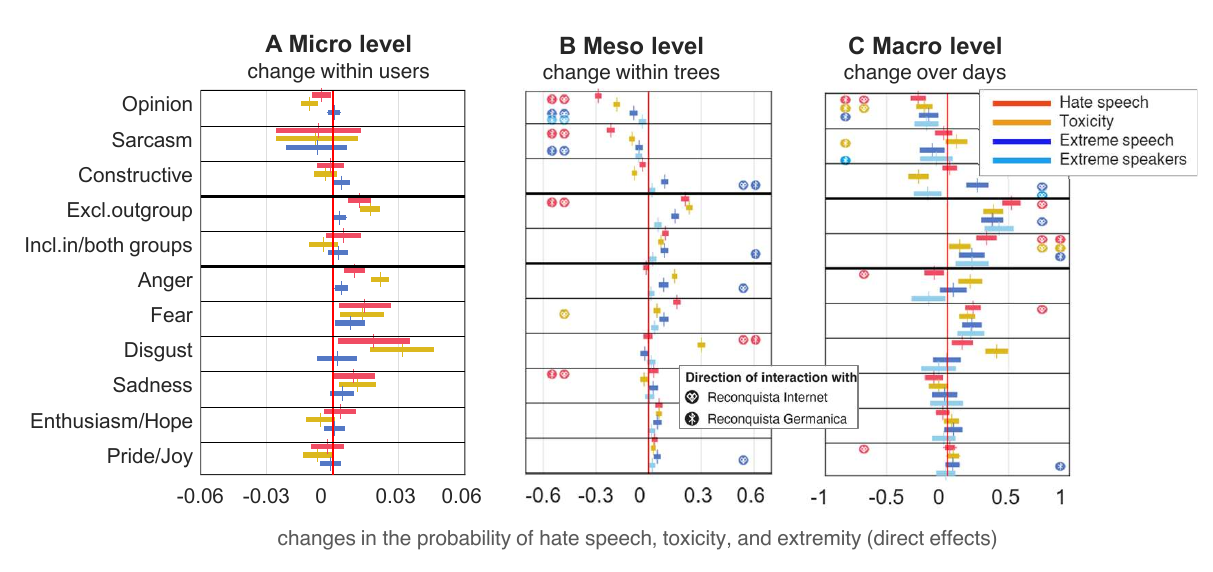}
    \vspace{0.2 cm}
    \caption{\footnotesize\textbf{Results of statistical models predicting changes in the probability of quality of discourse}. We predicted different indicators of the quality of discourse following tweets characterized by different dimensions of discourse. Positive coefficients mean that a dimension of discourse is related to an increase in hate speech, toxicity, and extremity of speech and speakers, while negative coefficients indicate a decrease. The left panel \textbf{(A)} shows the micro-level effects on a subsequent tweet, obtained via matching analysis. The middle panel \textbf{(B)} shows the direct meso-level effects within discussion trees, calculated as meta-analytic estimates from ARDL models fitted on 3,569 discussion trees. The right panel \textbf{(C)} shows the direct macro-level effects from day to day, obtained from ARDL models fitted on averaged dimensions of discourse over each of 1,461 subsequent days. The icons of Reconquista Germanica (combined letters R and X resembling a sword) and Reconquista Internet (a sign that resembles a heart) denote the direction of reliable interactions with the percentage of extreme speakers resembling one of the groups in each tree (panel B) and with the existence of one or both groups in the public sphere on a specific day (panel C). If an effect of a dimension became more negative (positive) when one or both of these groups were present, we add the respective icon to the left (right) side of the effect.  Additional results for lagged effects of discourse dimensions, robustness checks, and tables with all results, are provided in SI Sections~S10-S12.}
    \label{fig:ardl_results}
\end{figure}

At the micro level (Fig.~\ref{fig:ardl_results}A), we apply causal inference to estimate the effect of discourse dimensions in a reply tweet on the probability of hate speech in the next tweet written by the user who received the reply. We employ non-parametric matching to correct for confounding factors of the language of the tweet receiving the reply and of user and discussion characteristics, while also considering measurement error as identified in the validation of the classifiers used in our analysis. As a result, the coefficients obtained by our analysis are robust to limitations in the accuracy of the applied machine learning classifiers (see Methods for details). 

At the meso level (Fig.~\ref{fig:ardl_results}B), we use autoregressive distributed lag (ARDL) models to study the dynamics of discourse over successive tweets in 3,569 discussion trees that contained at least 50 tweets answering directly to the root tweet or to other tweets in the tree. To check the robustness of the tree-level results, we redid the same analyses only on 868 trees that contained at least 100 tweets, and on trees originating from different types of users --- large news organizations, individual politicians, or individual journalists and bloggers. Results reported in Section~S11 in the SI suggest that the patterns remain robust at different levels of analyses. 

Finally, at the macro level (Fig.~\ref{fig:ardl_results}C), we use ARDL models to study this dynamic over the 1,461 days in our time series (see Methods for details on both ARDL approaches). Although the users involved in the discourse might fluctuate, a day-to-day analysis can be informative about the community's general atmosphere and implicit rules about what can and cannot be said. Approximately half of the users in our data set tweeted more than once and these users are active for about a year on average (see Fig. S5 in the SI). The ARDL analyses on both meso- and macro levels help understand the effects of different dimensions of discourse on the subsequent discourse quality both directly on the very next tweet or the same day, and on up to two subsequent tweets or days. 

In all ARDL models, we also incorporate interaction effects of the dimensions of discourse with the presence of organized extreme groups, expecting that the presence of such groups creates a more polarized situation that could change the effectiveness of some strategies. On the meso-level, we compute the interaction of dimensions of discourse with the percentage of speakers resembling RI (counter speech group) or RG (hate speech group) in a tree. On the macro-level, we compute the interaction with the presence of either RI or RG at any of the given days. 

Overall, while there exist some differences between the results of the matching analysis and the ARDL models (as described next and in Sections S10-S12), the results for the different levels of discourse are broadly consistent. While the ARDL models cannot be used to derive causal interpretations, they account for autoregressive effects of the measures of quality of discourse on themselves, and the matching analyses allow for approximating causal effects in a setting where controlled experimentation is hardly feasible. 

\subsubsection*{Argumentation strategy}
Across all levels of analysis, offering simple opinions---not necessarily supported by facts, but without insults---is the most successful strategy for improving the quality of discourse. On the micro level, as shown in the first row of Fig.~\ref{fig:ardl_results}A, offering a simple opinion about an individual's tweet leads to a lower probability of hate and toxicity in the next tweet of the same individual. While these and other micro-level effects appear small, they are relatively large when compared to the baseline probability of hate speech in this corpus which is around 20\% (see Fig.~S6). On the meso level, as shown in the first row of Fig.~\ref{fig:ardl_results}B, the effect of the opinion strategy is again reliably negative for hate and toxicity, as well as for extremity of speech and speakers. Some of these effects are even more pronounced at times when discourse is dominated by organized groups Reconquista Germanica and Reconquista Internet, as indicated by the icons denoting interaction effects. On the macro level of discourse, as shown in the first row of Fig.~\ref{fig:ardl_results}C, offering opinions again reliably reduces hate, toxicity, and extremity. These effects are mostly immediate, showing in the very next tweet or on the same day, and typically not extending to the subsequent tweets or days (see SI Fig.~S19). The same holds for the other dimensions of discourse, so in what follows we comment only on their direct, immediate effects. Results for all lags can be seen in SI Figs. S19 and S20, and in Tabs. S13 and S14.  

Sarcasm can also be helpful, especially on the level of discussions where it consistently reduces hate, toxicity, and extremity of subsequent trees (second row of Fig.~\ref{fig:ardl_results}). Sarcasm also contributes to the reduction of extremity of speech over days. Note that here we use the umbrella term ``sarcasm'' for several distinct but empirically related styles of humor, including also irony, and cynicism~\cite{ruch2018}. Like the opinion strategy, sarcasm tends to be particularly useful in the presence of organized extreme groups. 

Constructive comments, such as providing facts, asking genuine questions, or exposing contradictions, have more mixed effects. It reduces the toxicity of discourse in subsequent tweets and days, and reduces the percentage of extreme speakers over days. However, it also consistently increases extremity of speech on all levels of analysis, especially in the presence of organized extreme groups (third row of panel Fig.~\ref{fig:ardl_results}). 

Taken together, the results so far indicate that in highly polarized situations, such as those when organized extreme groups compete for prevalence in discourse, offering simple opinions and using sarcasm might be more productive than striving to provide additional evidence and other constructive comments. This is in line with studies suggesting that receiving evidence that opposes one's strongly held beliefs can cause backlash~\cite{kahan2012polarizing, nyhan2014effective}). Note that the models do not include the insult strategy because it correlates highly with exclusionary statements about outgroups (see Method), described next. 

\subsubsection*{Ingroup/Outgroup content}
The results for ingroup or outgroup content in Fig.~\ref{fig:ardl_results} suggest that any mention of own or another group often relates to more hate, toxicity, and extremity. This is so not only when an outgroup is mentioned in an exclusionary way (the dominant way in which outgroups are mentioned in this corpus, see SI Fig.~S11), but also when both groups are mentioned in a more including context. This effect tends to intensify when one or both organized extreme groups are present in the discourse, possibly because it brings attention to and reinforces the already strong ingroup/outgroup divisions. 

\subsubsection*{Emotional tone}
In general, the results in the bottom half of Fig.~\ref{fig:ardl_results} show that negative emotions in tweets relate to more hate, toxicity, and extremity of the subsequent discourse, on micro-, meso-, and macro levels of discourse. This is true for anger as well, with one exception: especially when Reconquista Internet was active, anger contributed to less hate speech and fewer extreme users in the subsequent days. It is possible that the `righteous anger' of the Reconquista Internet members about the hate speech produced by Reconquista Germanica had temporarily discouraged some forms of hateful speech. At the same time, their anger could have attracted other commenters who felt the same~\cite{buerger2020anti}, possibly explaining the pronounced positive relationship between anger and extremity of speech on the meso level while RI was active.

Fear is reliably related to more hate, toxicity, and extremity at all levels of analysis. This is in line with findings that feelings of threat, especially threat from outgroups, contribute to heightened in-group identity and cohesion, possibly in preparation for defense~\cite{henderson1985cohesion,hogg2007uncertainty}. 

Disgust leads to more hate and toxicity at all levels of analyses. It does not affect extremity or might even lower the extremity of speakers over days. As a signal of inappropriateness of certain extreme positions, it might deter some of the extremists or motivate neutral users to participate more and express different views. 

Sadness has a more complicated relationship with the subsequent discourse. On the micro level, it worsens the quality of discourse. In contrast, on the meso- and especially on the macro levels it tends to slightly improve the quality of discourse. In line with \cite{garcia2019collective}, it is possible that some particularly sad external events such as terrorist attacks initially lead to a deteriorated or more negative discourse, but later prompt an increase in the overall level of solidarity in the community, improving the quality of discourse.

Positive emotions in tweets sent in reply to a particular user can contribute to somewhat lower toxicity of the subsequent tweets of that user, but on meso and macro levels they can---counterintuitively---lead to worse quality of discourse. It is possible that emotions such as enthusiasm, hope, pride, and joy are used primarily as a way to rally and unite one's own group, rather than to promote overall reconciliation and unity. This is supported by the fact that these effects are sometimes stronger in the presence of organized extreme groups. 

\section{Discussion}
\label{sec:discussion_nhb}

We analyzed discourse dynamics in a large corpus of Twitter discussions over four years. On the level of individual tweets, discussion trees, and days, we explored how different measures of discourse quality relate to the argumentation strategy, ingroup/outgroup content, and emotional tone of the preceding tweets, and how these relationships change in the presence of organized extreme groups. Our results, summarized in Fig.~\ref{fig:ardl_results} and detailed in SI Sections S10-S12, suggest that the most effective way of reducing hate, toxicity, and extremity of discourse in this corpus was to simply provide opinions, not necessarily supported by facts, and without insults. Sarcasm, irony, and cynicism were helpful as well, especially when used in the presence of organized extreme groups. More constructive comments, including providing facts and exposing contradictions, were helpful in reducing toxicity, but at the expense of increasing extremity of speech. It is possible that seeing evidence for one side of the argument caused backlash among some of the supporters of the other side, who found additional arguments for sticking to their initial position (c.f.~\cite{kahan2012polarizing,nyhan2014effective,avalle2024persistent}). 

Our results also suggest a strong role of both outgroup and ingroup content in fostering hate speech, toxicity, and extremity of discourse. While it is well-known that constructing and emphasizing boundaries between own and other groups is one of the most important aspects of social cognition~\cite{lamont2002study}, this dimension of discourse has not been explicitly measured in prior research on online hate and counter speech. Our finding that mentioning either own or other groups inflames tensions is striking, but it also has a flip side: discourse that does not explicitly mention social groups might, according to our results, diffuse hate, toxicity, and extremity, as the exchange between diverse groups of people is vital for a productive societal dialogue. The mentioning of group identities might make differences salient~\cite{hewstone2002intergroup} and lead to more intense discussions~\cite{avalle2024persistent}. Future studies could try to disentangle the exact circumstances under which ingroup and outgroup content is productive or detrimental.

When it comes to emotional tone,  messages expressing negative emotions such as anger, fear, disgust, and sadness generally increase the level of hate, toxicity, and extremity of the subsequent discourse. The exception are anger and sadness that can contribute to less hate, toxicity, and fewer extreme speakers in the following days. This is in line with previous studies finding that negative emotions such as anger generally lead to more negativity~\cite{paletz2023}. More suprisingly, while positive emotions such as enthusiasm and hope, pride and joy can diffuse toxicity, they can also lead to more subsequent hate and toxicity, possibly because they are often used to rally one's own group rather than to foster overall unity. 

Our conclusions are based on diverse measures of discourse dimensions, constructed using very different methods and training samples. We measure hate speech using a transformer-based language model fine-tuned to classify labeled examples from our corpus and validated against a held-out test set annotated by human raters. We show that our results are robust 
across different measures of discourse quality - hate speech, toxicity, extremity of speech and speakers - as well as on micro, meso, and macro levels of analysis. 

Our study of Twitter data provides valuable insights into the dynamics of online discourse. However, it is crucial to recognize that each social media platform fosters unique patterns of interaction and should be studied individually to understand platform-specific strategies. For example, Facebook, Instagram and TikTok are each characterized by long-form discussions within groups, high-engagement visual content, or fast-paced interactions, respectively. Understanding these differences is essential for comprehensively analyzing the varied landscape of social media discourse and its implications for public opinion and behavior. Still, each major social media platform provides variations of text-based comment and response functions, which speaks to the generalizability of our findings. Accordingly, past research has shown that dynamics across platforms are likely similar~\cite{avalle2024persistent}.

Our results are limited to the dimensions of discourse we investigated and are not informative about other kinds of misuse of online commons beyond hatefulness and extremism, such as various forms of misinformation and fraud. That should be a topic of further research. We could not make reliable conclusions about some counter speech strategies that have been identified in experimental studies, such as empathy for the victims of hate speech~\cite{munger2017tweetment,hangartner2021empathy} and related moral appeals to treat the outgroup well~\cite{munger2017experimentally}, because we found too few examples of such strategies in our corpus. It is possible that such an empathic emotional tone would also have promoted a more civil discourse. This raises the question of whether it is more productive to teach people new, potentially more effective strategies, or encourage strategies that people are already using even though they might be somewhat less effective. Furthermore, the relationship between the dimensions of discourse assessed in our study (argumentation strategy, outgroup content, and emotional tone) and the quality of discourse might be subject to confounding factors. For example, which beliefs are expressed can be influenced by more general social norms~\cite{alvarez2018normative}, and the motivations to express hate speech or not might lie outside of what happens on social media~\cite{davis1998klan}.

Of course, discourse aimed at public persuasion must always be conducted with an awareness of ethical implications and the potential for unintended consequences. Counter speech is no exception. Many organized counter speech groups adhere to strict codes of conduct. For example, \#iamhere ---currently the most well known and largest citizen organized group to fight hate and misinformation online---enforces rules such as: avoid spreading hate, prejudice, slander, gossip, or rumors; be respectful; stay factual and on topic; support others; and be inclusive. Despite their good intentions, counter speakers can still encounter unintended consequences. In the case of \#iamhere, the group was accused of bullying and brigading when, after a post was flagged for counter speech, so many members responded that they overwhelmed the conversation. This led to backlash against users associated with the \#iamhere hashtag~\cite{cathyiamhere}. At the same time, our analysis revealed that ethically questionable strategies such as insulting somebody, making the outgroup look weak, or responding with anger overwhelmingly lead to worse discourse quality, while more ethical strategies such as voicing opinions benefit the discourse as a whole.

Another question is whether citizens would be motivated to join collective moderation efforts on other social media platforms. Emotional arousal, which often accompanies uncivil discourse, can be related to increased engagement~\cite{bergerArousalIncreasesSocial2011} and citizens that aim to maximize other's engagement with content they produce might be less motivated to engage in moderation. However, there is also a growing body of literature suggesting that hateful discourse drives down user engagement~\cite{ziegele2016not, hickeyNoLoveHaters2023}. Historically, platforms that allow for unmoderated uncivil discourse, such as Gab, Parler, Truth Social, and since recently Twitter (now known as X), tend to draw less public engagement, have smaller user bases, and struggle to find advertisers willing to place their brand next to uncivil content. 
There is evidence of a widespread interest of citizens in performing collective civic moderation, as reflected in counter speech groups around the world. Currently, the most well-known and largest group is the iamhere international (\url{https://iamhereinternational.com/}) network, which includes 15 groups in North America, Asia and throughout Europe~\cite{cathyiamhere,buerger2020anti} and there are many other groups aiming to fight misinformation and hate in different countries. 
Overall, while some users wish to be uncivil, the community as a whole seems to be willing to engage in civil discourse with the aid of effective and fair civic content moderation. 

Our results provide a nuanced picture of the effect of different discourse dimensions on the quality of subsequent discourse at different time scales. As such, they can be useful to citizens and citizen groups who wish to tackle hate speech in their online spaces. Collective civic moderation can be effective for improving the quality of online discourse and managing the common-pool resources on social media platforms. 

The effectiveness of providing mere opinions and using a touch of sarcasm is particularly noteworthy, since it reduces the need for spending time on crafting nuanced arguments and lowers the barrier for citizens to engage in counter speech. Except for extreme positions, it is often a matter of how something is said instead of what has been said (e.g., ``Dear Miss Will I am very disappointed about your [...] moderation.'' instead of ``Lying media!'', Tab.~S2)~\cite{rosenberg2015nonviolent}. Opinions can benefit from I-messages~\cite{gordon2000parent} and following social media netiquette (e.g.,~\cite{ard-netiquette}) to avoid insults. Although rare in our corpus, people can try to focus on common ground and engage in decategorization~\cite{gaertner2000reducing}. We provide examples for all dimensions of discourse including beneficial and detrimental argumentation strategies in Tab. S2 in the SI.

Our results are encouraging for the possibility of both broader participation in formulating implicit norms of conduct, and easier monitoring of the discourse as it unfolds. Both factors are important for the successful management of online commons, as suggested by Ostrom~\cite{ostrom1990governing}. Also, in line with her observations that sanctions for breaking the rules need to be graduated, we find that strong negative reactions to others' discourse---for example, expressing negative emotions or using in/outgroup language---are often related to worse quality of the subsequent discourse. In sum, citizens should be educated and empowered to participate in online discussions by providing opinions (perhaps with a bit of sarcasm) without evoking strong emotions or provoking ingroup-outgroup divisions. Such speech could help improve the discourse in the long run even in the presence of organized extreme groups.

\backmatter

\section{Methods}
\label{sec:methods}

\begin{figure}[!ht]
    \centering
    \includegraphics[width=1\textwidth]{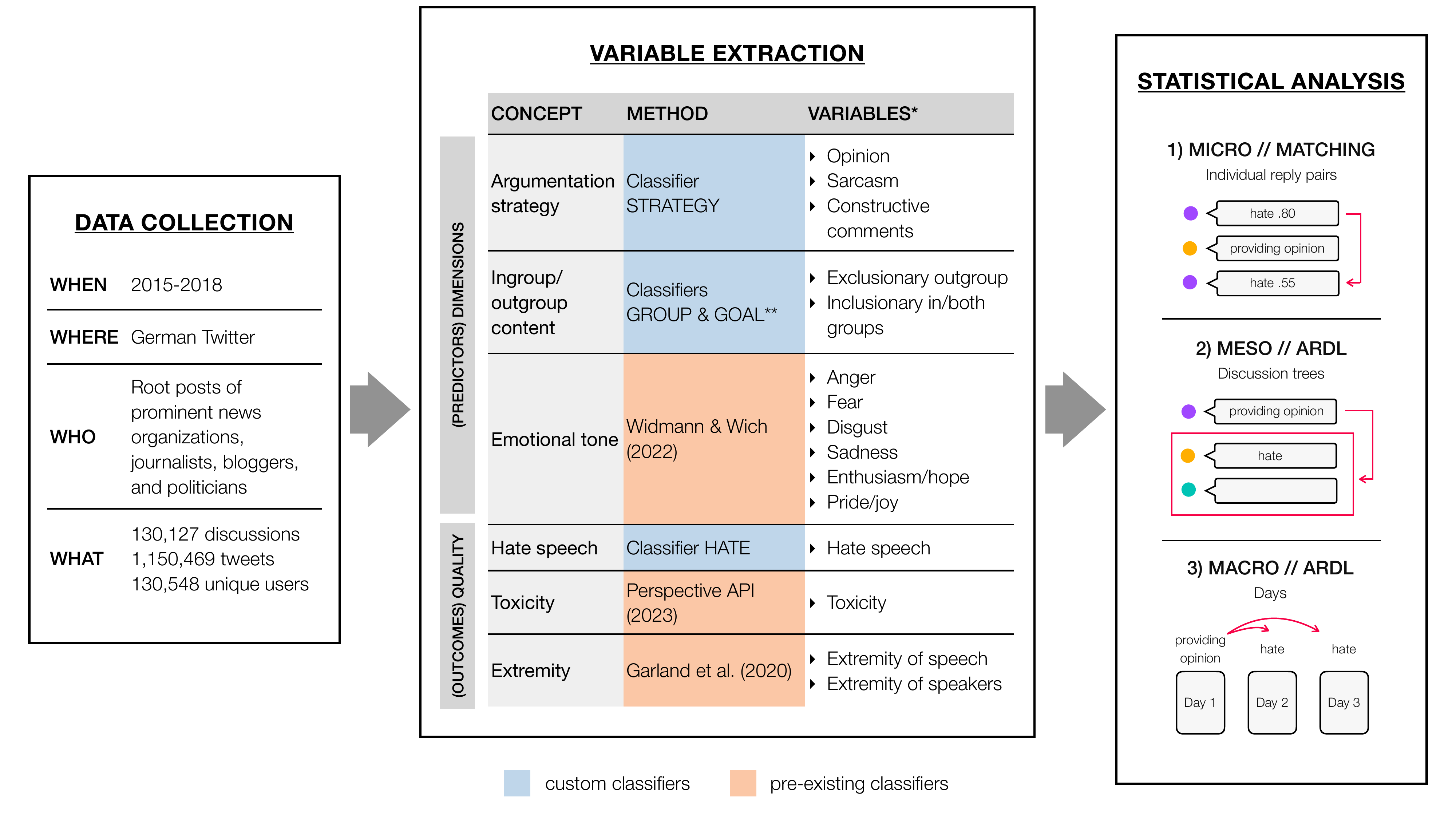}
    \caption{\footnotesize \textbf{Overview of the study}. Blue shading: We developed new classifiers to extract dimensions of discourse Argumentation strategy and Ingroup/outgroup content, as well as Hate speech, a measure of discourse quality. Orange shading: Where feasible, we applied pre-existing classifiers to detect discourse dimension Emotional tone and derive other measures of discourse quality - Toxicity and Extremity. We analyzed the relationship between dimensions of discourse and discourse quality on three different levels: 1) the micro level of individual reply pairs (numbers are examples for a tweet containing hate speech as measured by our classifier); 2) the meso level in the remainder of a discussion tree; and 3) the macro level over entire days. \\
    \textit{Notes}: *Column ``Variables'' lists the classes extracted by the classifiers that were used as predictors in the statistical analyses on all three levels. **Ingroup/outgroup content was extracted with two classifiers in conjunction: classifier GROUP identified whether in- and/or outgroup content was present at all, while classifier GOAL identified the socio-psychological goal of a tweet. Details are provided in the Methods.}    
    \label{fig:flowchart}
\end{figure}

Fig.~\ref{fig:flowchart} gives an overview of the different data processing and analysis steps, ranging from data collection and machine learning approaches to extract dimensions of discourse and indicators of discourse quality, to the different levels of statistical analysis. Details of data curation and training of the machine learning classifiers are also reported in the SI Section~S13, following the REFORMS checklist~\cite{kapoor2024reforms}.

\subsection{Data set}\label{sec:data}

We used a custom scraper to collect Twitter conversations or ``discussion trees'' originating from tweets posted by prominent German news organizations, journalists, bloggers, and politicians between January 2015 (the start of the so-called ``migrant crisis'' in Europe) and December 2018. At that time the Twitter API did not have the capacity to collect reply trees (conversations) in their entirety, which is what we needed for our analysis. As such, we developed a custom scraper system that would collect conversation URLs of interest and then parse the resulting HTML to reconstruct the reply trees from the raw HTML that was scraped from these URLs. In early 2019, Twitter made massive and abrupt changes to the way the conversation page HTML was generated with the express purpose of making it extraordinarily hard to scrape data from their website. As a result our data gathering process halted in early 2019.

This data set consists of 130,394 trees containing 1,167,853 tweets from 134,092 unique users. (Note that in \cite{garland2022impact} we had erroneously reported that we used 181,370 trees containing 1,222,240 tweets. The correct numbers are reported above.) After excluding tweets for which we were not able to calculate toxicity scores (see SI Section~S1.1 for details), we ended up with 130,127 trees containing 1,150,469 tweets from 130,548 unique users. 

It is important to acknowledge the challenges of collecting data---especially retrospectively---about posts that potentially violate platform rules. Such posts can be missing from the data due to content removal or user bans. While we captured data almost concurrently with its creation, ensuring a substantial collection of toxic speech, it remains possible that some violating content was missed. Notably, when we revisited the same time frame for additional data collection, much of the previously gathered content had been removed, and many associated users had been banned or suspended. This indicates that our initial data collection successfully captured a significant portion of the toxic speech before it was removed by the platform. For more details on the data set, its construction and limitations see~\cite{garland2020countering,garland2022impact} and Section S6 in the SI. 

\subsection{Measurement of discourse quality and dimensions of discourse}\label{sec:classification_scheme}
We use machine learning approaches to quantify both discourse quality and the dimensions of discourse. Discourse quality is operationalized by four disjunct measures: (1) the probability for a tweet to be hateful, (2) the probability for a tweet to be toxic, and (3) the similarity of a tweet to speech from known members of the organized groups Reconquista Germanica and Reconquista Internet (extremity of speech), and (4) the likelihood that a tweet was written by a member of those groups (extremity of speakers). Dimensions of discourse are (1) the argumentation strategy (opinion, sarcasm, constructive arguments, or insults), (2) whether tweets reference the author's ingroup or outgroups, and in which way (inclusionary or exclusionary), and (3) the emotional tone of the tweet. In addition, if we detect a tweet that is likely to contain hate speech, we also classify the target of the hate but do not use this information in the present article.

For political extremity, toxicity and emotional tone in tweets we use pre-existing classifiers. The classifier for political extremity classifies text on a scale from -1 to 1, with positive values indicating speech similar to RG and negative values indicating speech similar to RI~\cite{garland2020countering} (see also Section S1.1 in the SI). To classify toxicity, we use Google's Perspective API, which uses a multilingual BERT model to infer the toxicity of a tweet on a scale from 0 to 1~\cite{perspective-website} (see also Section S1.2 in the SI). For the inference of emotional tone, we use the classifier in ~\cite{widmann2022creating}, trained on a corpus of German tweets and outputting scores from 0 to 1 for each of anger, fear, disgust, sadness, as well as joy, enthusiasm, pride, and hope (see Section~\ref{subsec:classifying_emotions} for details). 

For hate speech, argumentation strategy, and ingroup/outgroup references we develop custom machine learning classifiers trained on human annotated data sets. To annotate the data sets, we develop a classification scheme for identifying hate, as well as for the dimensions of discourse capturing the argumentation strategy, in- and outgroup thinking, and the target of hate speech (see SI Section~S2 for details). Importantly, we make sure to balance our training data---including the data used to develop the classification scheme---over extremity of speech. This means that argumentation strategies were initially analyzed on a large portion of tweets similar to RI, a citizen-organized group aiming to counter hate in the online sphere, and as such thought to reflect counter speech. Furthermore, argumentation strategies even though not intended as counter speech in the first place can have counter speech effects.


We hand-annotate a considerable subset of tweets from the data set ($n=15,692$), following the classification scheme mentioned above. The labeling process including annotator training and interrater agreement is described in detail in SI Section~S3. Some characteristics of the labeling task (such as context dependency~\cite{arhin2021ground}) can drive down interrater agreement, without indicating low data quality as discussed in Supplement Section~3.7 and in~\cite{herderich2024measuring}. To make our results robust to suboptimal interrater agreement in this challenging labeling task, we propagate the classification error of our classifiers to the statistical analyses either by directly using class probabilities instead of binary classes in the ARDL models or by using agreement levels to sample the confidence intervals in the bootstrapping of the matching analysis. 

We use the labeled training data for supervised training of the three newly-developed machine learning classifiers (one for each of hate, argumentation strategy and ingroup/outgroup references). To this end, we use the pre-trained multilingual language model ``twitter-xlm-roberta-base''~\cite{barbieri2021xlm} which is based on XLM-R~\cite{conneau-etal-2020-unsupervised}, and which supports 100 different languages building on the RoBERTa~\cite{liu2019roberta} architecture.
RoBERTa is a deep learning model from the class of transformer models~\cite{vaswani2017attention}, which are currently used for a wide array of natural language processing tasks. Transformer models are designed to process sequential input---like textual data---to learn word embeddings that take a word's context into account. 
The model was pre-trained on a large corpus of multilingual text (including German) to learn general features of multiple languages. We then fine-tune it on the human annotated training data to learn the specific classification tasks. 
Details of the fine-tuning process are provided in the SI Section~S4.

\subsection{Validation of machine learning classifiers}\label{subsubsec:validation}
We validate the final classifiers using predicted and human labels on a held-out set of 1,000 examples that were labeled by all four of our annotators (see SI Section~S3.2 and Section~S5 for details). To create the ground-truth test set for validation of the classifiers, we only use labels on which at least three out of four annotators agreed. The final test set includes 900 examples for hate speech, 677 examples for argumentation strategy and 127 examples for in- and outgroup thinking. 

We develop five distinct models for each dimension of discourse (hate, argumentation strategy, and ingroup/outgroup references), trained on five different data splits from the last fine-tuning step (15 models in total, see SI Section~S4.2 for details). To calculate the final performance and performance variability, we use each of the models to predict the probability of belonging to a given class for every example in the held-out test set. 

We assess the receiver-operator characteristic~\cite{Hanley1982} of our classifiers and calculate the area under the curve (AUC)---a common metric to assess the performance of machine learning classifiers. ROC curves are calculated by comparing predicted class probabilities from each model (trained on different data splits) to human labels in the held-out test set. AUC values between 0.8 and 0.9 are typically considered as ``excellent discrimination'' while values between 0.7 and 0.8 are still considered as ``acceptable discrimination''~\cite{Hosmer2013}.
AUC values for our classifiers range from $0.73\pm 0.01$ (class ``exclusionary about the outgroup'' in in- and outgroup content) to $0.94\pm 0.01$ (class ``other'' in argumentation strategy). 
The AUC values for every class are reported in Tab.~S5 in the SI while the ROC-curves for every class are reported in Fig.~S3 in the SI.

We also transform the predicted class probabilities into class labels by assigning each example to the class with the maximum probability. 
We report the average classifier precision, recall and F1-score compared to a frequency-based random guessing benchmark for every class in Fig.~S4 in the SI. 

For the statistical analyses reported in Section~\ref{sec:results} we infer labels for all unlabeled tweets using the models with the best performance on the held-out test set. 
We note that for the statistical analysis (see Section~\ref{sec:statistical_analysis} below) we use the class probabilities that the models output directly, thereby mitigating inaccuracies from the decision to assign one label per tweet and dimension only, and incorporating uncertainty from the applied machine learning classifiers to make our estimates robust to measurement errors.

\subsection{Measurement of emotional tone}\label{subsec:classifying_emotions}
To classify the emotional tone of a tweet, we use a transformer-based classifier developed by Widmann and Wich~\cite{widmann2022creating}. The classifier detects eight discrete emotions---four positive (joy, enthusiasm, pride, hope) and four negative (anger, fear, disgust, sadness)---in text. We chose this classifier because it was specifically trained to measure affective language in German political speech, including text posted on social media~\cite{widmann2022creating}. As such, the data used to train the classifier is similar to the text contained in our corpus. The emotion detection classifier is based on a German version of ELECTRA~\cite{clark2020electra}, an extended version of BERT. 
The model reaches F1 scores of 0.60 (sadness, pride) to 0.84 (anger) on the test set and substantially outperforms other similar classifiers~\cite{widmann2022creating}. We use the classifier as published by Widmann and Wich to infer the probability to contain a given emotion for every tweet contained in our corpus. We note that the emotion classifier was trained on a multi-label classification task where every example can belong to more than one class. That is, for every example this classifier outputs a probability $p_i$ that the example contains the emotion $i$, but different to our custom trained machine learning classifiers, $\sum_i p_i >= 1$. For our statistical analysis, we again directly use the probabilities that are given by the model. 

\subsection{Statistical analyses}\label{sec:statistical_analysis}
We conduct analyses on three levels of discourse (for details see SI Section~S9). On the micro level, we use matching analysis to determine how tweets affect directly subsequent tweets, enabling us to understand what might happen if people used different discourse dimensions (argumentation strategies, ingroup/outgroup content, or different types of emotional tone; see SI Section S3.6 for example tweets in each category) to affect the content of direct replies to their tweets. Specifically, we identify discussion acts in which user A writes a tweet, is replied to by user B, followed by another tweet by user A in the same tree. This second tweet of user A could be in reply to user B or not - the only constraint is that there is no other tweet in the tree by user A after the reply written by user B. 

To approximate a random assignment of the strategy used in the tweet of user B, we apply non-parametric matching~\cite{hoMatchingNonparametricPreprocessing2007} to balance the treatment group with a subset of the control group for each discourse dimension (see SI Section S9 for details, and panel A in Fig.~\ref{fig:ardl_results} and SI Fig.~S18 for the results). The matching assures minimal difference between the treatment and the control groups with respect to a set of covariates that could bias the content of replies. After matching, we fit a linear regression model of the outcome score as a function of the treatment, including interaction effects and intercepts for all the covariates considered in the matching, i.e. double-adjusting~\cite{nguyen2017double} to correct for residual imbalances after matching. We calculate 95\% confidence intervals of causal effects via bootstrapping on the matched sample including a propagation of the classification error of the treatment into the uncertainty of the estimate (similarly to~\cite{card2022computational}). 

To investigate the relationships between the time series of discourse quality (hate speech, toxicity, and extremity of speech and speakers) and dimensions of preceding discourse (argumentation strategy, ingroup/outgroup content, and emotional tone), we used the autoregressive distributed lag (ARDL) modeling framework~\cite{kripfganz2018ardl,kripfganz2023ardl}, typically used for analyses of economic time series (see SI Section~S9.2). For our purposes, this framework is interesting because it enables estimation of the effects of the dependent variable on itself, as well as direct and lagged effects of independent variables, in a single-equation of the form:

\begin{equation}
    y_t=c_0+c_1  t+\sum_{i=1}^p \varphi_i y_{t-i} + \sum_{i=0}^q \beta_i x_{t-i}+u_t 
\end{equation} 

where $ c_0 $ is a constant, $ c_1 t $ is a time trend, $ y_{t-i} $ are lags of the dependent variable $ y_t $  with associated weights  $ \varphi_i $ denoting dynamic marginal effects of $ y $ on itself for $ p $ lags, $ x_{t-i} $ are lags of independent variables with the associated weights $ \beta_i $ denoting dynamic marginal effects of $ x $ on $ y $ for $ q $ lags, and where $ u_t $ is an error term. In this way, we can study each of the effects independently since weights for one variable are adjusted by the influence of other variables in the statistical model. As Dickey-Fuller tests show that our predictor variables are integrated of order 0 or 1 ($I(0)$ or $I(1)$) \cite{kripfganz2023ardl}, and all dependent variables of order 0, we do not explore the presence of cointegrating relationships with tests such as the bounds test~\cite{pesaran2001bounds} (see SI Section S9.2).

For each measure of discourse quality (hate speech, toxicity, extremity of speech and speakers), we apply ARDL models on two different levels of data aggregation. On the meso level, we aim to illuminate short-term dynamics within trees, by applying ARDL models over successive tweets in 3,569 discussion trees that contained at least 50 tweets (panel B in Fig.~\ref{fig:ardl_results} and SI Fig.~S19). To check whether longer discussion trees exhibit different discourse dynamics, we also replicate the analysis with 868 trees that include at least 100 or more tweets. Most patterns of results remain the same (Fig.~S21 in the SI), although some patterns are more pronounced when analyzing more (albeit shorter) trees. We could not analyze even shorter trees because our models would become overspecified.

On the macro level, we analyze more general and longer-term effects on the discourse quality, by applying ARDL models over average measures of each dimension for each of the 1,461 days in our time series (panel C in Fig.~\ref{fig:ardl_results} and SI Fig.~S20, and panel B in Fig.~S20). Note that these macro-level analyses include all tweets from all trees.

Because the number of predictors ARDL models can handle is limited (as each predictor is a complete time series), we could not include all potentially interesting predictors and their interactions. Specifically, we could not include the strategy of insults in the models, because of its high correlation with exclusionary references towards outgroups ($r=0.74$ on the level of tweets and $r=0.91$ on the level of days; see SI Tab.~S7 and Section~S8), leading to collinearity issues. Because insults are conceptually very similar to hate speech, one of the variables we wish to explain, we decided to focus on the conceptually more distinct category of exclusionary outgroup references instead. Furthermore, because one of our interests in this paper is exploring the effects of civic self-organization on discourse, we investigate how the presence of organized groups Reconquista Germanica (RG) and Reconquista Internet (RI) interacts with the effects of different discourse dimensions such as argumentation strategies. To enable a quick overview of these results in Fig.~\ref{fig:ardl_results} and SI Section S10, we mark the direction of all reliable interactions with the extremity of speakers for the tree-level analyses, and with the overall presence of RG and RI on Twitter for the day-to-day analyses using icons resembling the RG logo (a sign that combines letters R and X and resembles a sword) and the RI logo (a sign that resembles a heart), respectively. For example, if a dimension has an overall negative effect on hate speech, and the effect becomes even more negative when RG is active and/or present in a tree, then we add the icon for RG to the left of the effect. If the effect becomes more positive, we add this icon to the right of the effect; and we do the same for all reliable interactions with the presence of RI. All results are shown in detail in Section~S11 in the SI. For further details on all aspects of the statistical analyses, see SI Section~S9.

\clearpage

\section*{Acknowledgments}
We thank our annotators Lisa Trummer, Anna Spradley, and Elizabeth Michels for their valuable input for the classification scheme and their grit during the labeling process, Max Pellert for contributing scripts to interface with the emotion classification models, Cathy Burger, Henrik Olsson and Max Pellert for providing feedback on a draft version of this manuscript, and Liuhuaying Yang for her generous expert help with visualizations.

This manuscript was posted on a preprint: \url{https://arxiv.org/abs/2303.00357}

\section*{Declarations}
\subsection*{Funding}
JL was supported by the Marie Skłodowska-Curie grant No. 101026507. JL, AH, MG and the annotators were partially supported by the grant NSF DRMS 1757211. The annotators were also partially supported by an internal award from ASU's Center on Narrative, Disinformation and Strategic Influence. JG was partially supported by the Santa Fe Institute's Applied Complexity program. SA and DG were supported by the ERC Advanced Grant PRODEMINFO (101020961). MG was partially supported by the Austrian Research Promotion Agency grant number 873927 and the ERC Advanced Grant COLLADAPT (101140741).

\subsection{Competing interests}
The authors declare that they have no competing interests.

\subsection{Inclusion and ethics}
This study is based on publicly available archival Twitter data on German Twitter users and accounts of German major news outlets. This research activity is exempt from requiring IRB approval because the tweets are publicly available and used in a completely anonymized form (see paragraph 46.104-d-4 at \cite{IRB}).  

\subsection{Code \& data availability}
Following the Twitter terms of service, we are not allowed to publish the texts of the tweets contained in our corpus. We do, however, publish 
all inferred information necessary to reproduce the ARDL analysis presented in the paper. This data is available under accession code 10.17605/OSF.IO/X4WE6. 
The data can be provided by the authors pending scientific review and a completed material transfer agreement (see Supplementary). Requests for the data should be submitted to \url{alina.herderich@uni-graz.at} or \url{jana.lasser@uni-graz.at}.

Code for training the machine learning classifiers and conducting the statistical analyses is publicly available on GitHub: \url{https://github.com/JanaLasser/counterspeech-strategies}.

\subsection{Authors' contributions}
Authors' contributions according to the CRediT taxonomy distribute as follows: \\
Conceptualization: all authors \\
Methodology: JL, AH, JG, SA, DG, MG \\
Software: JL, JG \\
Validation: JL, AH, DG, MG \\
Formal Analysis: JL, AH, DG, MG \\
Investigation: JG, MG \\
Resources: JG \\
Data Curation: JL, AH, JG, MG \\
Writing -- Original Draft Preparation: JL, AH, JG, MG \\
Writing -- Review \& Editing: all authors \\
Visualization: JL, AH, DG, MG \\
Supervision: MG \\
Project Administration: JL, AH, MG \\
Funding Acquisition: JG, MG \\

Specifically JL developed the machine learning classifiers, AH developed the classification scheme and supervised the labeling process, JG provided the classifiers for extremity of speech and speakers, SA provided advice for the development of the machine learning classifiers, DG performed the matching analysis, and MG performed the ARDL statistical analyses.

\clearpage

\bibliography{references}
\nocite{krippendorff1970estimating}
\nocite{krippendorff2004}
\nocite{pavlopoulos2020}
\nocite{ross2017}
\nocite{mosqueira-reyHumanintheloopMachineLearning2022}

\end{document}